\documentclass[12pt,aps,cite,citesort,fullpage]{revtex4}

\usepackage{setspace}
\usepackage{apjfonts}
\usepackage{epsfig,graphicx}
\usepackage{pifont}
\usepackage{amssymb}
\usepackage{amsmath}
\usepackage{latexsym}
\usepackage{verbatim}
\usepackage{array}
\usepackage{xspace}
\usepackage{colordvi}
\usepackage{wasysym}
\usepackage{marvosym}

\newlength{\saveparindent}
\newcommand {\be}{\begin{equation}} 
\newcommand {\ee}{\end{equation}}

\def \be{\begin{equation}}
\def \ee{\end{equation}}
\def \bea{\begin{eqnarray}}
\def \eea{\end{eqnarray}}

\def \vo{v_o}

\def \xnu{x_{\nu}}

\def \xdot{\dot{x}}

\def \rdot{\dot{{\bf r}}}

\def \rddot{\ddot{{\bf r}}}

\def \Lag{\mathcal{L}}

\def \rA{{\bf r}_{\mbox{\tiny{A}}}}
\def \rB{{\bf r}_{\mbox{\tiny{B}}}}
\def \rBA{{\bf r}_{\mbox{\tiny{B/A}}}}
\def \vA{{\bf v}_{\mbox{\tiny{A}}}}
\def \vB{{\bf v}_{\mbox{\tiny{B}}}}

\def \pF{p_{\mbox{\tiny{F}}}}
\def \bfr{{\bf r}}

\def \r{{\bf r}}
\def \rstar{{\bf r}^\ast}
\def \ra{\bfr_{\mbox{\tiny A}}}
\def \rb{\bfr_{\mbox{\tiny B}}}
\def \pt{{\bf p}_t}
\def \ps{{\bf p}_s}
\def \rp{\bfr'}
\def \bfk{\boldsymbol{\kappa}}

\def \bfv{{\bf v}}
\def \vhat{\hat{{\bf v}}}
\def \vhatdot{\dot{\vhat}}

\def \ev{\hat{\mbox{{\bf e}}}_v}
\def \xhat{\hat{{\bf x}}}
\def \yhat{\hat{{\bf y}}}

\def \rhat{\hat{{\bf r}}}
\def \thetahat{\hat{\boldsymbol{\theta}}}
\def \sstar{s^{\ast}}
\def \curv{\boldsymbol{\kappa}}
\def \tangent{\hat{{\bf t}}}
\def \Dist{{\cal D}}
\def \Vk{V_{\kappa}}
\def \Vnc{V_{\mbox{\tiny NC}}}
\def \kc{\kappa_{\mbox{\tiny C}}}

\def \w{\omega}

\def \D{\partial}

\def \dij{\delta_{ij}}

\def \eps{\epsilon}

\def \d{\delta}

\def \dij{\delta_{ij}}

\def \l{\lambda}

\newcounter{saveeqn}%
\newcommand{\alpheqn}{\setcounter{saveeqn}{\value{equation}}%
\stepcounter{saveeqn}\setcounter{equation}{0}%
\renewcommand{\theequation}{\mbox{\arabic{saveeqn}\alph{equation}}}}%
\newcommand{\reseteqn}{\setcounter{equation}{\value{saveeqn}}%
\renewcommand{\theequation}{\arabic{equation}}}%

\begin{document}

%

\pagenumbering{arabic}
\setcounter{page}{1}

\title{Generalization of distance to higher dimensional objects}

\author{Steven  S. Plotkin$^\dag$\footnote{e-mail: steve@physics.ubc.ca}
}

\affiliation{ 
$^\dag$ Department of Physics and Astronomy, University of British
Columbia, 6224 Agricultural Road, Vancouver, BC V6T1Z1, Canada
}
\newpage
\vspace{10cm}
\newpage

\baselineskip=0.2cm


\begin{abstract}
The measurement of distance between two objects is generalized to the case where the objects are no longer points but are one-dimensional. Additional concepts such as non-extensibility, curvature constraints, 
and non-crossing become central to the notion of distance. Analytical and numerical results are given 
for some specific examples, and applications to biopolymers are discussed. 
\end{abstract}

\maketitle
\normalsize

\section{Introduction}

The distance, as conventionally defined between two zero-dimensional
objects (points) $A$ and $B$ at positions ${\bf  r}_{\mbox{\tiny{A}}}$ and ${\bf
  r}_{\mbox{\tiny{B}}}$, is the minimal arclength travelled in
the transformation from $A$ to $B$. A transformation ${\bf r} (t)$
between $A$ and $B$ is a vector function which may be
parametrized by a scalar variable $t$: $0\leq t\leq T$, ${\bf r} (0) = {\bf
  r}_{\mbox{\tiny{A}}}$, ${\bf r} (T) = {\bf
  r}_{\mbox{\tiny{B}}}$, and the distance travelled is a functional of
${\bf r} (t)$.  The (minimal) transformation ${\bf r}^\ast (t)$ is an object of
dimension one higher than $A$ or $B$, i.e. it yields a distance that is one-dimensional. The
distance $\mathcal{D}^\ast$ is found through the variation of the functional~\cite{GelfandIM00}:
\alpheqn
\bea
\mathcal{D}^\ast &=& \mathcal{D}\left[{\bf r}^\ast (t)\right] \mbox{where ${\bf r}^\ast (t)$
  satisfies} \\
&& \d \int_0^T \!\!\! dt \: \left(g_{\mu\nu} \dot{x}^{\mu}(t)
\dot{x}^{\nu}(t) \right)^{1/2} = 0 \: . \label{1dg} \\
\mbox{or} && \d \int_0^T \!\!\! dt \: \sqrt{\rdot^2} = 0 \;\;\;\; \mbox{(Euclidean metric)}
\label{1d}
\eea
\reseteqn
Here $\dot{x} = dx/dt$, and $\rdot=d\bfr/dt$. 
The boundary conditions mentioned above 
are present at the end points of the integral. 
The Einstein summation
convention will be used where convenient, e.g. eq.~(\ref{1dg}), however all the analysis here deals
with spatial coordinates, $\nu = 1,2,3$ on a Euclidean metric. Generalizations to 
dimension higher than $3$, as well as non-Euclidean metrics, 
are straightforward to incorporate into the formalism. 

On a Euclidean metric, $g_{\mu\nu}=\d_{\mu\nu}$ and the minimal distance
becomes the diagonal of a hypercube. However, formulated as above, the 
solutions minimizing $\mathcal{D}$ are infinitely degenerate, because particles
moving at various speeds but tracing the same trajectory over the
total time $T$ all give the same distance. To circumvent this problem
what is typically done is to let one of the space variables (e.g $x$) become the
independent variable. However for higher dimensional objects, or zero
dimensional objects on a manifold with nontrivial topology, there is
no guarantee that the dependent variables ($y$, $z$) constitute single valued
functions of $x$.  
Alternatively, one can study the 'time' trajectory of the
parametric curve defined above, but under a gauge that fixes the speed
to a constant $\vo$, for example. One can either fix the gauge from the outset with Lagrange 
multipliers, or choose a gauge that may simplify the problem after finding the extremum equations. 
The latter is often simpler in practice. 

To be specific, the effective Lagrangian $\Lag$ appearing in the above
problem is $\sqrt{\rdot^2}$, and the Euler-Lagrange (EL) equations are
\be
\frac{d}{d t}\left(\frac{\D \Lag}{\D \rdot}\right) = 0 \;\;\;\;\;
\mbox{or} \;\;\;\;\; \frac{d}{d t}\left(\frac{\rdot}{\left| \rdot \right|}\right) = \dot{\vhat} = 0
\label{eompt}
\ee
with $\vhat$ the unit vector in the direction of the velocity. The boundary conditions are
\be	
\bfr^\ast(0) = \rA \;\;\; \mbox{and} \;\;\; \bfr^\ast(T) = \rB \; .
\label{bc1d}
\ee

Since the derivative of a unit vector is always orthogonal to that vector, equation~(\ref{eompt}) says that
 the direction of the velocity cannot change, and therefore straight line motion results. Applying the
 boundary conditions gives $\vhat = (\rB-\rA)/\left|\rB-\rA\right|$. However, {\it any} function $\bfv (t) = \left|
 \vo(t)\right| \vhat$ satisfying the boundary conditions is a solution, so long as $\int_0^T \!\! dt \: \left|
 \vo(t)\right| = \left|\rB-\rA\right|$. This is the infinite degeneracy of solutions mentioned above. Then
 $\bfr^\ast(t) = \rA + \frac{\rB-\rA}{\left| \rB-\rA\right|} \int_0^t \!\! dt \: \left| \vo(t)\right|$, and
 $\Dist^\ast = \int_0^T\!\! dt \: \sqrt{\rdot^{\ast^2}} =  \int_0^T \!\! dt \: \left| \vo(t)\right| =
 \left|\rB-\rA\right|$. At this point we could fix the parameterization by choosing $\left| \vo(t)\right| =
\left|\rB-\rA\right|/T$ (constant speed), for example. 

The extremum is a minimum, as can be
shown by analyzing the eigenvalues of the matrix $\D^2 \mathcal{D}/\D \xnu(t) \D
x_\mu(t') = -\d_{\mu\nu}
\d''(t-t')$. Diagonalizing by Fourier transform gives 
positive elements $+ \omega_n^2 \,
\d_{\mu\nu} \d(\omega_n-\omega_n')$ for the stability matrix and thus
positive eigenvalues.  

In what follows we generalize the notion of distance to higher
dimensional objects, specifically space-curves. We will see many of the above themes reiterated, as
well as some fundamentally new features that emerge when one treats the space curves as non-extensible, 
having some persistence length or curvature constraint, 
and non-crossing or unable to pass through themselves. 
We provide analytical and numerical results for some prototypical examples for non-extensible chains, and 
we lay the foundations for treating curvature and non-crossing constraints.

\section{Distance metric for one dimensional objects}
\label{sec:polymer}

The distance $\Dist^\ast$ between two one-dimensional objects (which we refer to as space curves or strings) $A$ and $B$ having configurations 
$\ra(s)$ and $\rb(s)$, $0\leq s\leq L$, is obtained from the transformation from $A$ to $B$ that minimizes the
 integrated
distance travelled. By integrated distance we mean the cumulative arclength all elements of the string had to move in the transformation from $A$ to $B$. For the transformation to exist, strings $A$ and $B$ must have the same length (although this condition may be relaxed by allowing specific extensions or contractions). For the distance to be finite, open space curves must be finite in length. For closed 
non-crossing space curves, $A$ and $B$ must be in the same topological class for the transformation to exist. 
Describing the transformation $\bfr (t,s)$ requires two scalar parameters, one for arc length $s$ along the string and another 
measuring progress as in the zero-dimensional case, say $t$: $0\leq t\leq T$, so that $\bfr(s,0)=\ra(s)$ and 
$\bfr(s,T)=\rb(s)$. The distance travelled is a functional of the vector function $\bfr(s,t)$. The minimal 
transformation $\bfr^\ast(t,s)$ is an object of dimension one higher than $A$ or $B$, i.e. it yields a 
distance that is two-dimensional. The problem does not map to a simple soap film, since there are many 
configuration pairs that have zero area between them but nonzero distance travelled, e.g. a straight line 
displaced along its own axis, or that in figure~\ref{curvepairs}C. The analogue to a 
higher-dimensional surface of minimal area when the 'time' $t$ is included is closer but inexact (see footnote
below). 

We can construct the effective Lagrangian along the same lines as the zero-dimensional case. Using the 
shorthand $\bfr \equiv \bfr(s,t)$, $\rdot \equiv \D \bfr/\D t$, $\bfr' \equiv \D \bfr/\D s$, the distance 
travelled is${}^{\footnotemark}$ \footnotetext{The distance-metric action in eq.~(\ref{dsc}) bears a strong resemblance to the Nambu-Goto action for a 
classical relativistic string~\cite{ZwiebachB04}: $S_{\mbox{{\tiny NG}}} [\bfr (s,t)] = \int \!\! d\sigma \, 
d\tau \: \sqrt{(\rdot \cdot \bfr')^2 - (\rdot)^2 (\bfr')^2 }$, where $\bfr$ in $S_{\mbox{{\tiny NG}}}$ is 
now a four-vector and the dot product is the relativistic dot product. This action is physically interpreted 
as the (Lorentz Invariant) world-sheet area of the string. If eq.~(\ref{dsc})
could be mapped by suitable choice of gauge to the minimization of the Nambu-Goto action, one
could exploit here the same reparameterization invariance that results in wave equation solutions to 
the equations of motion for the classical relativistic string, by choosing a 
parameterization such that $\rdot \cdot\bfr' =0$ (for the purely geometrical problem, the discriminant under 
the square root in the action has opposite sign). Unfortunately however, because the velocity in the 
distance-metric action is a $3$-velocity rather than a $4$-velocity, our action only accumulates area when 
parts of the string move in $3$-space, in contrast to the Nambu-Goto action which accumulates area even for a 
static string. The distance-metric action eq.~(\ref{dsc}) has a lower symmetry than that for the classical 
relativistic string. $\Dist^\ast$ cannot depend on the time the transformation took, while the world sheet area does. Conversely, if we take e.g. configuration $A$ at $t=0$ to be a straight line of length $L$, and configurations $B$ at $t=T$ to be the same straight line but displaced along its own axis by varying amounts $d$, the geometrical area for all transformations would be $LT$, while the distances $\Dist^\ast_{\mbox{{\tiny AB}}}$ for each transformation would be $Ld$.}
\be
\mathcal{D}= \int_{0}^L\!\!\! ds \!\! \int_0^T \!\!\! dt \: \sqrt{\dot{x}^{\nu} \dot{x}_{\nu}} = 
\int_{0}^L\!\!\! ds \!\! \int_0^T \!\!\! dt \: \sqrt{\rdot^2} \: .
\label{dsc}
\ee

However to meaningfully represent the distance a string must move to reconfigure itself from conformation $A$ to $B$, the transformation must be subject to several auxiliary conditions. 

The first of these is non-extensibility. Points along the space curve cannot move independently of one another but are constrained to integrate to fixed length, so the curve cannot stretch or contract. Thus there is a Lagrange multiplier $\lambda(s,t)$ weighting the (non-holonomic) constraint:
\be
\sqrt{\bfr'^2} =1 \: .
\label{constraintL}
\ee
This constraint ensures a parameterization of the string with unit tangent vector $\tangent=\bfr'$, so that 
the total length of the string is $L = \int_{0}^L\!\! ds \: \sqrt{\bfr'^2} = \int_{0}^L\!\! ds$. In the language of differential geometry, the space curve is a unit-speed curve.

If the constraint~(\ref{constraintL}) were not present in eq.~(\ref{dsc}), each point along the space-curve 
could follow a straight line path from $A$ to $B$ and the problem of minimizing the distance would be trivial. 
Equivalently, setting $\lambda=0$ should reduce the problem to a sum of straight lines analogously to the 
zero-dimensional case above.

As in the case of distance between points, one can fix the $t$-parameterization from the outset by introducing 
a Lagrange multiplier $\alpha(t)$ that 
fixes the total distance covered per time $\int_{0}^L\!\! ds \sqrt{\rdot^2}$ to a known function $f(t)$.
While this approach removes the infinite degeneracy 
mentioned above, as a global isoperimetric condition it reduces the symmetry of the problem. For example there 
would then be no conservation law that could be written to capture the invariance of the effective Lagrangian
with respect to the independent variable $t$. For these reasons we choose to leave the answer as unparamaterized with respect to $t$, analogous to the point-distance case above. 

\subsection{Ideal chains}
\label{sec:gc}

There are many examples of nontrivial transformations between two strings $A$ and $B$ where chain non-crossing is unimportant (c.f. figures~\ref{curvepairs}A and~\ref{curvepairs}B).
Here we derive the Euler-Lagrange equations for this case.

From equations~(\ref{dsc})-(\ref{constraintL}), the extrema of the distance $\Dist$ are found from
\bea
\d \Dist = &\d& \!\!\! \int_0^L\!\! \int_0^T \!\!\! ds \, dt \: \mathcal{L}\left(\rdot,\bfr'\right) = 0 
\nonumber \\
\mbox{or} \;\;\; &\d& \!\!\! \int_0^L\!\! \int_0^T \!\!\! ds \, dt \: \left( \sqrt{\rdot^2}-\l \sqrt{\bfr'^2}\right) = 0
\label{DL}
\eea
Performing the variation gives
\bea
\d \Dist &=& \int_0^L\!\!\! ds \: \left[ \pt \cdot \d\r \right]_0^T + 
\int_0^T\!\!\! dt \: \left[ \ps \cdot \d\r \right]_0^L \nonumber \\
&-& \int_0^L\!\! \int_0^T \!\!\! ds \, dt \: \d \r \cdot \left[ \frac{d\ps}{d s}
+ \frac{d\pt}{d t}\right] =0
\label{dD}
\eea
where the generalized momenta $\pt$ and $\ps$ are given by:
\be
\pt = \frac{\D \Lag}{\D \rdot} = \vhat \;\;\;
\mbox{and} \;\;\; \ps =  \frac{\D \Lag}{\D \rp} = -\l \tangent
\ee
where $\vhat$ is again the unit velocity vector, and $\tangent$ is the unit tangent to the curve. 

The EL equation follows from the last term in~(\ref{dD}), and yields a partial differential equation for the minimal transformation $\rstar(s,t)$:
\be
\left( \rdot^2\right) \rddot - \left(\rdot \cdot \rddot \right) \rdot  =  \left| \rdot\right|^3 \left(  \lambda \bfr'' + \lambda' \bfr' \right)
\label{de-ghost}
\ee
where we have used the facts that $\left| \bfr'\right|=1$ and $\bfr'\cdot\bfr''\equiv \hat{{\bf t}}\cdot \boldsymbol{\kappa} =0$, since the tangent is always orthogonal to the curvature at any given point along a space curve. 

Equation~(\ref{de-ghost}) can be written in terms easier to understand intuitively by using the unit velocity vector $\vhat$, tangent $\tangent$, and 
curvature $\curv$:$^{\footnotemark}$\footnotetext{The invariance of the Lagrangian to $(s,t)$ leads to conservation laws by Noether's theorem~\cite{GelfandIM00}, which here take the form of divergence conditions. However these generally contain no new information beyond the EL equations, and can be obtained by dotting eq.~(\ref{de-EL}) with either
$\bfr'$ to give $\l' = \vhatdot \cdot \tangent$, or $\rdot$ to give $\bfv \cdot (\l \tangent)' =0$.}
\be
\dot{\vhat} = \l \curv + \l' \tangent \: .
\label{de-EL}
\ee

Comparison of equations~(\ref{de-EL}) and~(\ref{eompt}) illustrates the point made earlier that setting 
the Lagrange multiplier $\l$ corresponding to the non-extensibility condition to zero results in straight line 
solutions for all points along the space curve. Conversely the condition that the space curve form a 
contiguous object results generally in nonzero deviation from straight line motion.
So in comparing various extremal solutions to eq.~(\ref{de-EL}), the minimal solution will minimize $\left| \l \right|$ everywhere.

The boundary conditions are obtained from the first two terms in~(\ref{dD}). Since the initial and final configurations are specified, the variation $\d\r$ vanishes at $t=0,T$, and the corresponding boundary conditions, or initial and final conditions, are:
\be
\rstar(s,0) = \rA(s) \;\;\;\;
\mbox{and} \;\;\;\; \rstar(s,T) = \rB(s)  \: .
\ee

Since the end points of the string are free during the transformation, $\d\r \neq 0$ at $s=0,L$, and so the conjugate momenta must vanish: $\ps(0,t) = \ps(L,t) = 0$. This means that $\l\tangent = 0$ at the end points. However since $\tangent$ cannot be zero, the only way this can occur is for $\l(0,t)=\l(L,t)=0$. The Lagrange multiplier, which represents the conjugate force or tension to ensure an inextensible chain, must vanish at the end points of the string. If $\l=0$, the EL equation~(\ref{de-EL}) gives $\dot{\vhat} = \l' \tangent$ at the end points. However since $\vhat$ is a unit vector, $\dot{\vhat}$ is orthogonal to $\vhat$ (or $\bfv$), and we have finally the boundary conditions at the end points of the string:
\be	
\l' \bfv \cdot \tangent =0 \;\;\;\;\; \mbox{(at the end points).}
\label{bcs}
\ee

Equation~(\ref{bcs}) has three possible solutions. One is that $\bfv \cdot \tangent =0$ or equivalently 
$\rdot \cdot \rp =0$, which corresponds to {\it pure rotation} of the end points. It is worth mentioning that the end points of the classical relativistic string also move transversely to the string. Moreover because of the Minkowski metric the end points must also move at the speed of light. Here however because Lorentz invariance is not at issue, additional solutions are possible. The end-points of our string can be at rest, $\bfv =0$, and satisfy the boundary condition~(\ref{bcs}). The last solution of eq.~(\ref{bcs}) is for $\l'=0$. Because $\l$ also vanishes at the end points, eq.~(\ref{de-EL}) gives $\dot{\vhat} =0$, or {\it straight line motion}. In summary the three possible boundary conditions for the string end points are:
\alpheqn
\bea
\bfv \cdot \tangent &=& 0 \;\;\;  \mbox{(pure rotation)} \label{rotbc} \\
\bfv &=& 0 \;\;\; \mbox{(at rest)} \label{restbc} \\
\dot{\vhat} &=& 0 \;\;\; \mbox{(straight line motion)}
\label{stlinebc}
\eea
\reseteqn

Whether an extremal transformation is a minimum can be determined by 
examining the second variation of the functional~(\ref{DL}):
\be
\d^2 \Dist = \frac{1}{2} \int_0^L\!\! \int_0^T \!\!\! ds \, dt \: \left[
  \d\rdot \cdot \mathbf{I} \cdot \d\rdot
+ \d\rp \cdot \boldsymbol{\Lambda} \cdot \d\rp \right] \: ,
\label{2nd}
\ee
where $\mathbf{I}_{ij} =  (\rdot^2 \dij - \xdot_i \xdot_j)/|\rdot|^3$ and $\boldsymbol{\Lambda}_{ij} = -\l(s,t) \, \dij$, and $\d\rp$ and $\d \rdot$ are the $s$ and $t$ derivatives of the variation $\d\bfr$ from the extremal path.

We now apply these concepts to some specific examples.

\subsection[]{Examples}

{\bf Translations.} If two space curves differ by a translation, $\rB(s)=\rA(s)  +{\bf d}$ with ${\bf d}$ a constant vector. The appropriate boundary condition for the end points is~(\ref{stlinebc}). The points along the string can all satisfy~(\ref{de-EL}) with $\dot{\vhat} =0$ and $\l=0$ everywhere (since $\tangent$, $\curv \neq 0$), and straight line motion results: $\rstar(s,t) = \rA(s) + (\rB(s)-\rA(s))t/T$. The distance $\Dist^\ast = L \left| {\bf d} \right|$.  This is the 1-dimensional analogue to eq.s~(\ref{eompt}),~(\ref{bc1d}).

{\bf Piece-wise linear space curves.}
Suppose initially the curvature of some section of the string is zero. Then, taking the dot product of $\bfv$ with eq.~(\ref{de-EL}), we see that eq.~(\ref{bcs}) holds for {\it all} points along the string. So the string either rotates or translates (or remains at rest if that segment has completed the transformation). 

Generally if one string partner has curvature (e.g. $\rA$ in fig.~\ref{curvepairs}B) the transformation is more complicated, but if both $\rA$ and $\rB$ are straight lines as in figure~{\ref{curvepairs}A, equation~(\ref{bcs}) holds for both. It is then reasonable to seek solutions $\rstar$ of the EL equation such that equation~(\ref{bcs}) holds for all (s,t). 

Consider the two space curves shown in figure~\ref{curvepairs}A with $\ra(s)=s\, \xhat$ and $\rb(s)=s\, \yhat$, both with curvature $\bfk=0$. We first investigate rotation from $A$ to $B$. This transformation satisfies the EL equation so appears to be extremal: $\r = s \rhat = s (\cos \w t \xhat +  \sin\w t \yhat)$.
The velocity $\rdot = s\w \thetahat$, so the Distance $\Dist [\bfr_{\mbox{\tiny{ROT}}} (s,t)] = \pi L^2/4$.
Taking the dot product of $\tangent$ with eq.~\ref{de-EL} gives $\l' = \tangent \cdot \vhatdot = -\w$, or 
$\l (s,t) = \l_o -\w s$. For the transformation to be extremal, the conjugate momenta must also vanish at the string end points, or $\l(0,t) = \l(L,t)=0$. This is impossible to achieve with this functional form, so the transformation is not extremal. 

We may however include the subsidiary condition here that $\rA(0,t)=\rB(0,t)$. Then the end point of the
 string at $s=0$ is determined, and the variations $\d \bfr(0,t)$ must vanish. Now only $\l(L,t)=0$, and so
 $\l(s,t) = \w (L-s)$. The transformation is extremal. 

Whether it is a minimum can be determined by 
examining the second variation~(\ref{2nd}). For the transformation $\bfr_{\mbox{\tiny{ROT}}} (s,t)$, 
the matrix $\mathbf{I}$ in~(\ref{2nd}) is non-negative definite, a necessary condition for a local minimum~\cite{GelfandIM00}, 
however $\boldsymbol{\Lambda}$ is negative definite, so the character of the extremum is determined by the 
interplay of the two terms in~(\ref{2nd}). Variations $\d\bfr$ that preserve $\rp^2=1$ or $2 \tangent \cdot 
\d\rp=0$ are satisfied in this example by $\d\bfr = f(s,t) \thetahat$, where $f(s,t)$ must satisfy the boundary 
conditions $\d\bfr(0,t)=\d\bfr(s,0)=\d\bfr(s,T)=0$. We thus let the variations have the functional form: 
$\d\bfr = \eps \sin(k s) \sin(n\pi 
t/T) \thetahat$, where $\thetahat = -\sin \w t \xhat +\cos \w t \yhat$, $n$ is a positive integer, and $k$ is 
unrestricted. Inserting this functional form for the variations into eq.~(\ref{2nd}) gives $\d^2 \Dist =
(\eps^2 \pi/8) \mathcal{F} (k L)$, where $\mathcal{F} (x)$ is a non-positive, monotonically decreasing 
function, with a maximum of zero at $kL=0$. In fact to lowest order $\mathcal{F} (k L) \approx -( \pi 
\eps^2/2160) \, (k L)^6$. The extremum corresponding to pure rotation of curve $\rA$ into $\rB$ is a maximum!

The only other solution to equations~(\ref{de-EL}) and~(\ref{bcs}) for all $(s,t)$ is for each point $s$ on $\ra(s)$ to be connected to a corresponding point on $\rb(s)$ by a straight line, corresponding to equation~(\ref{stlinebc}). Equation~(\ref{bcs}) holds everywhere because $\l'(s,t)=0$. Because $\l$ is zero at the boundaries it is thus zero everywhere. 

An intermediate configuration then has the shape of a piecewise linear curve with a right angle 'kink' at $\sstar(t)$ (see fig~\ref{L-curve}). As $t$ progresses, the kink propagates along curve $\rb$, and the horizontal part of the chain follows straight line diagonal motion, shrinking as its left end is overlaid onto curve $\rb$. The solution for the velocity at all $(s,t)$ is given by 
$ \bfv(s,t) = v_o (t) \Theta\left(s- \sstar(t)\right) \ev $
where $\sstar(t)$ is the position of the tangent discontinuity in figure~\ref{L-curve}, which goes from 
$\sstar(0)=0$ to $\sstar(T)=L$ as $t$ goes from $0$ to $T$. $\ev$ is 
a unit vector along the direction of the velocity, $\ev=(-\xhat+\yhat)/\sqrt{2}$, and $v_o(t)$ is a speed
which can be taken constant. By simple geometry, $v_o = \sqrt{2}\, \dot{s}^{\, \ast}$. Because $s^{\, \ast}(T)=L$, $v_o = \sqrt{2}L/T$ and $s^{\,\ast}(t) = L\, t/T$. The total distance travelled from equation~(\ref{dsc}) is then $\Dist^{\ast} = L^2/\sqrt{2}$. 

Because the transformation involves straight line 
motion, it is minimal. This can be seen from the second variation eq.~(\ref{2nd}). 
The shape of the curve at all times is given by
\bea
\bfr^\ast (s,t) &=& s \,\, \Theta (Lt/T-s ) \, \yhat  + (Lt/T) \Theta (s-Lt/T ) \, \yhat \nonumber \\
&+& (s-Lt/T) \Theta (s-Lt/T ) \, \xhat 
\label{r-L2}
\eea
Taking variations from the extremal path as before, let $\d\bfr = \eps \sin k (s-Lt/T) \sin(n\pi 
t/T) \Theta (s-Lt/T) \yhat$. These variations only act on the ``free'' part of the string and preserve a unit tangent to first order. The matrix $\boldsymbol{\Lambda}$ in~(\ref{2nd}) is zero for straight line transformations where $\l=0$.
The quadratic form $\d\rdot \cdot \mathbf{I} \cdot \d\rdot$ is non-negative, and results in a 2nd variation $\d^2\Dist = \eps^2 (32 \sqrt{2})^{-1} [ (kL)^2 + (n \pi)^2 (1 - \mbox{sinc}^2(k L) )]$, which is non-negative,  monotonically
increasing in $kL$, and quadratic to lowest order, with a minimum of zero at $kL=0$. The transformation is indeed minimal. 

Likewise, the minimal distance to fold a string of total length $L$ upon itself starting from a straight line (to form a hairpin) is $\Dist^\ast = L^2/4$.

{\bf Solution Degeneracy.} The above example illustrates that there are essentially an infinite number of extremal transformations: 
one can piece together various rotations and translations for parts or all of the chain while
still satisfying the EL equations. This infinity of extrema is likely to lead to nearly insurmountable
difficulties for the solution of eq.~(\ref{de-ghost}) by direct numerical integration. For these 
reasons we apply a method based on analytic geometry to obtain numerical solutions. This described in 
more detail below. 

There is also an infinite degeneracy of solutions having the minimal distance in the above example. To see a second minimal transformation, imagine running the above solution backwards in time, so the kink propagates from $s=L$ to $s=0$ along $\rb$. But this solution should hold forwards in time for the original problem if we permute $\rb$ and $\ra$. Now intermediate states $\rstar$ first run along $\xhat$, then $\yhat$. But then we can introduce multiple right angle kinks in various places, without causing the trajectories in the transformation to deviate from straight lines, so that intermediate states look like staircases. As there are an infinite number of possible staircases in the continuum limit, there is an infinite degeneracy. 
This can lead to a tangent vector $\rp$ whose magnitude is length-scale dependent, and less than unity until 
$s\rightarrow 0$. For example an 
intermediate configuration can be drawn in figure~\ref{L-curve} which appears as a straight diagonal line from 
$\rstar(0,t)$ to $\rstar(L,t)$, until $s\rightarrow 0$ when an infinite number of step discontinuities are 
revealed. This problem is resolved in practice through finite-size effects involving different critical angles of rotation described below. In the continuum limit it is resolved by introducing curvature constraints.

{\bf Curvature constraints.} In applications to polymer physics, chains have a stiffness characterized by bending potential in the analysis that is proportional to the square of the local curvature. Here we may choose to characterize stiffness by introducing a constraint on the configurations of the space curve, so that the curvature simply cannot exceed a given number:
\be
\Vk \left(\bfr''\right) = \Theta\left(\left| \bfr''\right| < \kc \right) \: .
\label{Vcurv}
\ee
This term lifts the infinite degeneracy mentioned above, as each near-kink (with putative $\kappa > \kc$) 
would result in slight deviations from linear motion in the above example, and thus an additional cost in the 
effective action. Other functional forms for $\Vk$ are also possible. For some applications a more conventional stiffness potential of the form 
$ \Vk \left(\bfr''\right) = \frac{1}{2} A_{\kappa} \bfr''^2  $
may be more appropriate.  However then the action would no longer consist of a true distance functional, and 
its minimization would involve the detailed interplay of the parameter $A_{\kappa}$ favouring globally minimal 
curvature with other factors affecting distance in the problem. 

{\bf Discrete Chains.} Strings with a finite number of elements (chains) provide a more accurate 
representation of real-world systems such as biopolymers. Discretization is also essential for numerical solutions in these more realistic cases. Monomers on a discretized chain travel along a curved metric~\cite{GrosbergAY04}, and Lagrange multipliers explicitly account for this fact here.

We start by discretizing the string into a chain of $N$ links each with length $ds=L/N$, so that 
equation~(\ref{dsc}) becomes $(ds) \int \! dt \, \sum_{i=1}^{N+1} \sqrt{\rdot_i^2}$, with each $\bfr_i(t)$ a 
function of $t$ only. 
The total distance is the accumulated distance of all the points joining the links, plus that of the end points, all times $ds$. 
This approach is essentially the method of lines for solving equation~\ref{de-EL}: 
the PDE becomes a set of $N+1$ coupled ODEs. 

Equation~(\ref{constraintL}) becomes $N$ constraint equations added to the effective Lagrangian: 
$\sum_{i=1}^{N} \hat{\lambda}_{i,i+1} \sqrt{(\bfr_{i+1}-\bfr_{i})^2}$. 
We rewrite this strictly for convenience
as $\sum \frac{\l_{i,i+1}}{2} \: \bfr_{i+1/i}^2$, where $\bfr_{i+1/i} \equiv \bfr_{i+1} - \bfr_i$, and 
$| \bfr_{i+1/i} | = L/N$.

The PDE in~(\ref{de-EL}) then becomes $N+1$ coupled (vector) ODEs, each of the form
\be 
\vhatdot_i + \l_{i-1,i} \, \bfr_{i/i-1} - \l_{i,i+1} \, \bfr_{i+1/i} = 0
\label{de-discrete}
\ee
with $\l_{0,1}=\l_{N+1,N+2} = 0$. Equation~(\ref{de-discrete}) is consistent with~(\ref{de-EL}) after suitable
definitions, for example the curvature at point $i$ after discretization is given by 
$(\bfr_{i+1/i} - \bfr_{i/i-1})/ds^2$.

{\bf One link.} We turn to the simplest problem of one link with end points $A$ and $B$ (see fig.~\ref{fig1link}), for which the action reads 
$L \int_0^T \! dt\,  ( \sqrt{\rdot_{\mbox{{\tiny A}}}^2} + \sqrt{\rdot_{\mbox{{\tiny B}}}^2} - \frac{\l(t)}{2} \rBA^2 )$. Points $A$ and $B$ have boundary conditions $\rA(0)=\bf{A}$, $\rB(0)=\bf{B}$, 
$\rA(T)=\bf{A'}$, $\rB(T)=\bf{B'}$. The link in our problem is taken to have a direction, so point $A$ cannot
transform to point $B$. The Euler-Lagrange equations become:
\begin{gather}
\begin{matrix} \vhatdot_{\mbox{{\tiny A}}} - \l \, \rBA = 0 \\ 
\vhatdot_{\mbox{{\tiny B}}} + \l \, \rBA = 0  \end{matrix} \quad 
\text{or} \quad
\begin{matrix} \l \, \vA \cdot \rBA = 0 \\
\l \, \vB \cdot \rBA = 0 \end{matrix}
\label{de-link}
\end{gather}
where the orthogonality of $\bf{v}$ and $\vhatdot$ has been used. 

Reminiscent of eq.~(\ref{bcs}), equations~(\ref{de-link}) each have $3$ solutions. For point $A$ these are:
(1) $\vA \cdot \rBA = 0$, or {\it pure rotation} of $A$ about $B$, 
(2) $\vA =0$ or point $A$ is {\it stationary}, or 
(3) $\l = 0$ and thus $\vhatdot_{\mbox{{\tiny A}}} = 0$ from the EL equations, indicating {\it straight-line
motion}. 
Moreover, (1) implies $\vB = 0$, or both points rotate about a common center, (2) implies $\vB \cdot \rBA = 0$ or $B$ rotates, 
and (3) implies $\vhatdot_{\mbox{{\tiny B}}} = 0$ as well, so that {\it both} points move in straight lines. 
An extremal transformation thus involves either straight line motion, or rotations of one point about the other at rest (or common center).  Once again, there are an infinite number of solutions: any combination of translations and rotations satisfies the EL equations, such as those shown in figure~\ref{fig1link}B-F. 

The Lagrange multiplier may be found from the first integral: taking the dot product of 
the EL equation for $B$ with $\rBA$ gives $-ds^2 \l = \rBA \cdot \vhatdot_{\mbox{{\tiny B}}}$. 
Thus when $B$ moves 
in a straight line $\l=0$. When $B$ rotates about $A$, its acceleration ${\bf a}_{\mbox{{\tiny B}}}$ follows
from rigid body kinematics as ${\bf a}_{\mbox{{\tiny A}}} + \boldsymbol{\alpha} \times \rBA - \omega^2 \rBA$,
where $\boldsymbol{\omega}$ and $\boldsymbol{\alpha}$ are the angular velocity and acceleration respectively, and ${\bf a}_{\mbox{{\tiny A}}}=0$. Thus $\l = 1/L$.

The minimal solution is the one that involves the minimal amount of rotation (and monotonic approach to $A'B'$). This may be obtained from 
analytic geometry: for the example configurations in fig.~\ref{fig1link}F, point $B$ rotates about point $A$ 
until $B''$, where the straight line $\overline{B'' B'}$ is tangent to the circle of radius $ds=L$ about $A$. 
The distance (over $ds$) is $AA' + L \theta_c + B'' B'$, where 
$\sin \theta_c = L/(L+A A')$ and $B'' B' = \sqrt{(A A')^2 + 2 L (A A')}$, so for example
if $A A' = 2 L$, $\Dist \approx 5.168\, L^2$.

{\bf Chains with curvature.}
We can now investigate the transformation shown in figure~\ref{curvepairs}B with the above methods. This is 
the canonical example when at least one of the space curves has non-zero curvature $\curv$.
Let $\ra = R \sin (\pi s/2 L) \xhat + R \cos (\pi s/2 L) \yhat$ and $\rb = s \xhat + R \yhat$, with $0\leq s\leq L$ and $R=2 L/\pi$.
We then discretize the chain into $N$ segments. 
According to eq.~(\ref{de-discrete}), the end point 
velocities $\vhatdot_{1}$, $\vhatdot_{N+1}$ obey EL equations of the same form as equations~(\ref{de-link}), and thus either rotate or translate. 
The situation for these links is analogous to figures~\ref{fig1link}B and~\ref{fig1link}F, in that 
the angle the link must rotate depends on the order of translation and rotation. 
The geometry in figure~\ref{curvepairs}B is analogous to transformations $A'B' \rightarrow AB$ in 
figures~\ref{fig1link}B, \ref{fig1link}F, in that the critical angle $\theta_c$ the link must rotate before 
translating is smaller if translation occurs first. 

Figure~\ref{curvchaintot} shows the two minimal solutions thus obtained. The transformation in 
fig.~\ref{curvchaintot}A undergoes translation away from curve $\rA$, and rotation at $\rB$. It is the global minimum. The transformation in~\ref{curvchaintot}B rotates from $\rA$ through a larger critical angle (see~\ref{curvchaintot}B inset), and then translates to $\rB$. Both solutions have a soliton-like kink 
that propagates across either space-curve $\rB$ or $\rA$. 

The minimal transformation follows these steps: (1) Link $\bfr_{2/1}$ rotates about $\bfr_1$, $\bfv_1=0$, 
$\bfv_2 \cdot \bfr_{2/1}=0$, and the Lagrange multiplier representing the conjugate 'force' $\l_{12}\neq 0$. 
During this rotation, nodes $3,4, \ldots$ move in straight lines formed by their initial values 
$\bfr_{\mbox{{\tiny A}}3}, \bfr_{\mbox{{\tiny A}}4}, \ldots$ and the tangent points to circles of radius $ds$ 
centered at $\bfr_{\mbox{{\tiny B}}2}, \bfr_{\mbox{{\tiny B}}3}, \ldots$. The corresponding Lagrange 
constraint forces $\l_{23}, \l_{34}, \ldots$ are all zero. 
Links $\bfr_{3/2}, \bfr_{4/3}, \ldots$ all adjust their orientation to ensure straight-line motion of their 
end points (dashed lines in fig.~\ref{curvchaintot}A), except for $\bfr_2$ which follows a curved path. 
(2) When link $\bfr_{2/1}$ completes its rotation, it coincides with curve $\rB$, and the process starts again 
with link $\bfr_{3/2}$ which begins its rotation about $\bfr_2$, while nodes $4, 5, \ldots $ move in straight 
lines. This process continues until the final link $\bfr_{N+1/N}$ rotates into place on $\rB$. 
The transformation in~\ref{curvchaintot}B is essentially the time-reverse of the above, but starting at  
curve $\rB$ and ending on $\rA$. 

For ideal chains without curvature constraints, the distances obtained from the two transformations in~\ref{curvchaintot}A,B differ non-extensively as the number of links $N\rightarrow \infty$. Moreover, the distance for each transformation itself differs non-extensively from the Mean Root Square distance 
$MRSD = N^{-1} \sum_{i=1}^{N} \sqrt{(\bfr_{\mbox{{\tiny A}}i} -\bfr_{\mbox{{\tiny B}}i})^2 }$ as $N\rightarrow 
\infty$.~$^{\footnotemark}$\footnotetext{The MRSD is always less than or equal to the Root Mean Square Deviation or RMSD between structures, as can be shown by applying H\"{o}lder's inequality.}
Specifically, the distance travelled by straight line motion scales as $ds \, NL \sim L^2$, while 
the distance travelled by rotational motion scales as $ds \, (N \overline{\theta}_c ds) \sim L^2/N$. 

On the other hand, curvature constraints as in eq.~(\ref{Vcurv}) become more severe on consecutive links
as $N\rightarrow \infty$, and can yield extensive corrections to the distance. Specifically, the increase
in distance $\Delta \Dist$ due to curvature constraints scales like the radius of curvature $R$ times $N$, since every node is affected by the rounded kink as it propagates. So $\Delta\Dist \sim ds \, N R \sim L R$. 
The importance of this effect then depends on how $R$ compares to $L$ (the ratio of the persistence length to the total length). It does not vanish as $N\rightarrow \infty$. Non-crossing constraints described below also 
yield extensive corrections to the distance travelled. 

\subsection{Non-crossing space curves}
\label{sec:ncsc}

The minimal transformation may be qualitatively different when chain crossing is explicitly disallowed.  Figure~\ref{curvepairs}C illustrates a pair of curves that differ only by the order of chain crossing. They are displaced in the figure for easier visualization but should be imagined to overlap so the quantity $\int_0^L \left| \ra -\rb\right| \approx 0$, i.e. if they were ghost chains their distance would be nearly zero, and most existing metrics give zero distance between these curve pairs (see Table~I).

Analogous to the construction of Alexander polynomials for knots, if we form the orthogonal projection of these space curves onto a plane there will be double points indicating one part of the curve crossing over or under another.  To transform from configuration $\ra$ to $\rb$ without crossing, the curves must always go through configurations having zero double points. If we trace the curve in an arbitrary but fixed direction, each double point occurs twice, once as underpass and once as an overpass. We may call the part of the curve between two consecutive passes a bridge. If the bridge ends in an overpass we assign it +1, if the bridge ends in an underpass we assign it -1, so traversing from the left in figure~\ref{curvepairs}C, curve $\rb$ has (+1) sense, and curve $\ra$ (-1). The change in sense during any transformation obeying non-crossing is always $\pm 1$, while ghost chains can have changes of $\pm 2$. 

The non-crossing condition means that the Lagrangian for the minimal transformation now depends on the position $\bfr(s,t)$ of the space curve, which may be accounted for using an Edwards potential:
$
\Vnc ([\bfr(s,t)] ) = \int_0^L \!\!\! ds_1 \!\!\! \int_0^L \!\! ds_2 \:\, \d ( \bfr(s_1,t) - \bfr(s_2,t) ) 
$
In practice a Gaussian may be used to approximate the delta function, with a variance that may be adjusted to account for the thickness or volume of the chain.

The Euler-Lagrange equation now becomes 
\be
(\Vnc)_{\bfr} = (\mathcal{L}_{\rp})_s + (\mathcal{L}_{\rdot})_t - [(\Vk)_{\bfr''}]_{ss}
\label{nc-EL}
\ee
where the curvature potential in eq.~(\ref{Vcurv}) has been included, 
and the notation $(\mathcal{L}_{\rp})_s \equiv (d/ds)(\D \mathcal{L}/\D \rp)$ has been used.
Equation~(\ref{de-EL}) is now modified to 
\be
\vhat_t =  \left( \l \tangent \right)_s + \nabla \Vnc + [(\Vk)_{\bfr''}]_{ss} 
\ee

To access various conformations, the minimal transformation must now abide by the non-trivial geometrical constraints that are induced by non-crossing. In general this renders the problem difficult, however the 
example in figure~\ref{curvepairs}C is simple enough to propose a mechanism for the minimal transformation consistent with the developments above, without explicitly solving the EL equations in this case. 
In analogy with the hairpin transformation described below eq.~(\ref{r-L2}), the transformation here
involves essentially forming and then unforming a hairpin. 
$\rA(N)$ (the blue end of curve $\rA$ in fig~\ref{curvepairs}C) propagates back along its own length until it reaches the junction, where it then rotates over it to become the overpass (this takes essentially zero distance in the continuum limit). The curve then doubles back following its path in reverse to its starting point. This transformation is fully consistent with the allowed extremal rotations and translations of the discretized chain. The distance in the continuum limit is $\Dist = \int_0^\ell \!\! ds \, (2 s) = \ell^2$, where $\ell$ is the length of the shorter arm extending from the junction in fig~\ref{curvepairs}C.

\section{Discussion}

The distance between finite objects of any dimension $d$ is a variational problem, and may be calculated by minimizing a vector functional of $d+1$ independent variables. Here we formulated the problem for space curves, where the function $\bfr^{\ast}(s,t)$ defining the transformation from curve $\ra$ to curve $\rb$ gives the minimal distance $\mathcal{D}$. 

We provided a general recipe for the solution to the problem through the calculus of variations. 
For simple cases the solution is analytically tractable. Generally there are an infinity of extrema, 
and direct numerical methods are unlikely to be fruitful. We employed a method that interpreted the 
discretized EL equations geometrically to obtain minimal solutions. The various solutions obtained here are 
summarized in Table~I, and compared with other similarity measures currently used. 

The distance metric may be generalized to higher dimensional manifolds, for example a two dimensional surface needs three independent parameters to describe the transformation. The distance becomes $\mathcal{D} = \int \! du \int \! dv \int \! dt \, |\rdot|$ and the constant unit area condition becomes $\left| \frac{\D \bfr}{\D u}\times \frac{\D \bfr}{\D v}\right| =1$. 

The question of a distance metric between configurations of a biopolymer has occupied the minds of many in the protein folding community for some time (c.f. for example~\cite{Leopold92,Chan94,FalicovA96,DuR98:jcp,ChoSS06}). Such a metric is of interest for comparison between folded structures, as well as to quantify how close an unfolded or partly folded structure is to the native. Chan and Dill~\cite{Chan94} investigated the minimum number of moves necessary to transform one lattice structure to another, in particular while breaking the smallest number of hydrogen bonds. Leopold {\it et al}~\cite{Leopold92} investigated the minimum number of monomers that had to be moved to transform one compact conformation to another. Falicov and Cohen investigated structural comparison by rotation and translation until the minimal area surface by triangulation was obtained between two potentially dissimilar protein structures~\cite{FalicovA96}. 

The present theoretical framework allows computation of a minimal distance between proteins of the same length by rotating and translating until $\mathcal{D}$ is minimized, as done in the calculation of RMSD. Comparison between different length proteins would involve the further optimization with respect to insertion or deletion of protein chain segments. 

It is interesting to ask which folded structures have the largest, or smallest average distance $\left< \mathcal{D} \right>$ from an ensemble of random coil structures, and also whether the accessibility of these structures in terms of $\mathcal{D}$ translates to their folding rates. It can also be determined whether the distance to a structure correlates with kinetic proximity in terms of its probability $\pF$ to fold before unfolding~\cite{DuR98:jcp}, by calculating $\left< \mathcal{D} \pF \right>$. 
The question of the most accessible or least accessible structure may be formulated variationally as a free-boundary or variable end-point problem. 

It is an important future question to address whether the entropy of paths to a particular structure is as important as the minimal distance. In this sense it may be the finite "temperature" ($\beta < \infty$) partition function $Z(\beta ) = \int d[\bfr(s,t)] \exp\left(- \beta \mathcal{D}[\bfr(s,t)]\right)$, i.e. the sum over paths weighted by their 'actions', which is the most important quantity in determining the accessibility between structures. This has an analogue to the quantum string: we investigated only $Z(\infty)$ here. 
We hope that this work proves useful in laying the foundations for unambiguously defining distance between biomolecular structures in particular and high-dimensional objects in general.

\section{Acknowledgements}

We are grateful to Ali Mohazab, Moshe Schecter, Matt Choptuik, and Bill Unruh for insightful discussions. 
Support from the Natural Sciences and Engineering Research Council and the A. P. Sloan Foundation is gratefully acknowledged.



\setcounter{section}{0}
\renewcommand{\theequation}{\arabic{equation}}%



\newpage

\centerline{FIGURE CAPTIONS}

FIGURE \ref{curvepairs}:
Three representative pairs of curves. {\it A} Straight line curve rotated by $\pi/2$.
{\it B} One string has a finite radius of curvature, the other is straight. {\it C} A canonical example where
non-crossing is important- the curves are displaced for easy visualization but should be imagined to be 
superimposed.

\vspace{0.3in}

FIGURE \ref{L-curve}:
The minimal transformation from {\it A} to {\it B} in figure~\ref{curvepairs}A involves the propagation of a kink along curve {\it B}. The end point of the curve at intermediate states satisfies $x+y=L$, the equation for a straight line. A similar linear equation holds for any point on the curve, thus no solution with shorter distance can exist. An intermediate configuration is shown in red. Alternative transformations are possible with kinks along {\it A}, as well as multiple kinks (see text).

\vspace{0.3in}

FIGURE \ref{fig1link}:
Transformations between two rigid rods. (A) undergoes simultaneous translation and rotation and so is not
extremal. (B) is extremal and minimal. The rod cannot rotate any less given that it translates first. However 
this transformation is a weak or local minimum.  (C), (D), and (E) are extremal but not minimal. (F) Is the global minimum. It rotates the minimal amount, and both $A$ and $B$ move monotonically towards $A'$, $B'$. 
A purely straight-line transformation exists but involves moving point $A$ away from $A'$ before moving towards it (similar to (D)), thus covering a larger distance than the minimal transformation. 

\vspace{0.3in}

FIGURE \ref{curvchaintot}:
Two minimal transformations between the curves shown in fig.~\ref{curvepairs}B, for $N=10$ links. 
Fig (A) is the global minimal 
transformation $\bfr^\ast(s,t)$, with $\Dist^\ast \approx 0.330 \, L^2$, figure (B) is a local minimum
with $\Dist \approx 0.335 \, L^2$.
In~(A), links with one end touching curve $\rB$ rotate, the others translate first from $\rA$, rotating only 
when one end of a link has touched $\rB$. In (B) they rotate first from $\rA$, then translate into $\rB$. 
Dashed lines in (A) show the paths travelled for each bead. The inset of (A) plots the total distance 
travelled as a function of the number of links $N$, with various $N$ plotted as filled circles to indicate the 
rapid decrease and asymptotic limit to $\Dist_{\infty} \approx 0.251 \, L^2$
The inset in~(B) shows the minimal angle each link must rotate during the transformation- it is less for the 
transformation in (A). 
Movie animations of these transformations are provided as Supporting Information. 

\newpage

\vspace{0.5in}
\centerline{TABLES AND TABLE CAPTIONS}
\vspace{1in}
\begin{table}[h]
\begin{ruledtabular}
\caption{Values of the distance for various examples considered here, compared to other metrics.}
\begin{tabular}{l|c|c|c|c}
{Curve Pair} & {$\Dist^\ast \, (L^2)$} & {\footnotesize RMSD}${}^\star \,$ {\footnotesize (L)} & {\footnotesize (1-Q)}${}^\dagger$ & {$\chi^{\sharp}$}  \\
\hline
Trivial translation & $|{\bf d}|/L$ & $|{\bf d}|/L$ &  0 & 0 \\
``L-curves'', fig~\ref{curvepairs}A &  $1/\sqrt{2}$ &  $\sqrt{2/3}$ &  --${}^\ddag$ & 0 \\
Straight line to Hairpin  & $1/4$ & $1/\sqrt{6}$  & 1 & 1/2 \\
``C-curve''- st. line, fig~ \ref{curvchaintot}A & $0.330$ & 0.371 & --${}^\ddag$  & 0.417 \\
``C-curve''- st. line, fig~ \ref{curvepairs}A${}^\natural$ & $0.251 $ & 
$0.334$ &
--${}^\ddag$  & 1 \\
``Over/under'' curves, fig~\ref{curvepairs}C & $(\ell/L)^2$  & $\approx 0$ &  $0^\wr$ & 0 \\
Single link, fig~\ref{fig1link}F${}^{\flat}$ & $5.168$ & $\sqrt{7}^\lambda$  & --${}^\delta$ & --${}^\delta$ \\

\end{tabular}
\end{ruledtabular} 
\begin{tabular}{l}
${}^\star$ {\footnotesize{$ RMSD\equiv \surd{N^{-1} \sum_i 
(\bfr_{\mbox{{\tiny A}} i}- \bfr_{\mbox{{\tiny B}} i})^2}$}} 
${}^\dagger$ {\footnotesize{Fraction of shared contacts $A$ has with $B$,}} \\
{\footnotesize see~\cite{DuR98:jcp,ChoSS06} for definitions.} \\
${}^{\sharp}$  {\footnotesize{Structural overlap function equal to $1$ minus the fraction of residue pairs}} \\  {\footnotesize{with similar distances in structures $A$ and $B$. The formula in ref.~\cite{VeitshansT96} is used.}}  \\
${}^{\natural}$  {\footnotesize{i.e. In the continuum limit.}}  
${}^{\flat}$  {\footnotesize{For $AA' = 2 \times \mbox{link length}$.}}
${}^\ddag$ {\footnotesize{$0/0$ or undefined}} \\
${}^\wr $ {\footnotesize{Assuming a contact is made at the junction.}}
${}^\delta$ {\footnotesize{Undefined for a single link}} \\
${}^\lambda${\footnotesize $\Dist$ is larger than the RMSD here because RMSD contains a factor of $2$
while $\Dist$} \\ 
{\footnotesize did not. We could have computed the ``effective distance'' for the rod by dividing} \\
{\footnotesize by $2$.}

\end{tabular}
\label{table1}
\end{table}

\newpage

\begin{figure}
\includegraphics[width=0.412\linewidth]{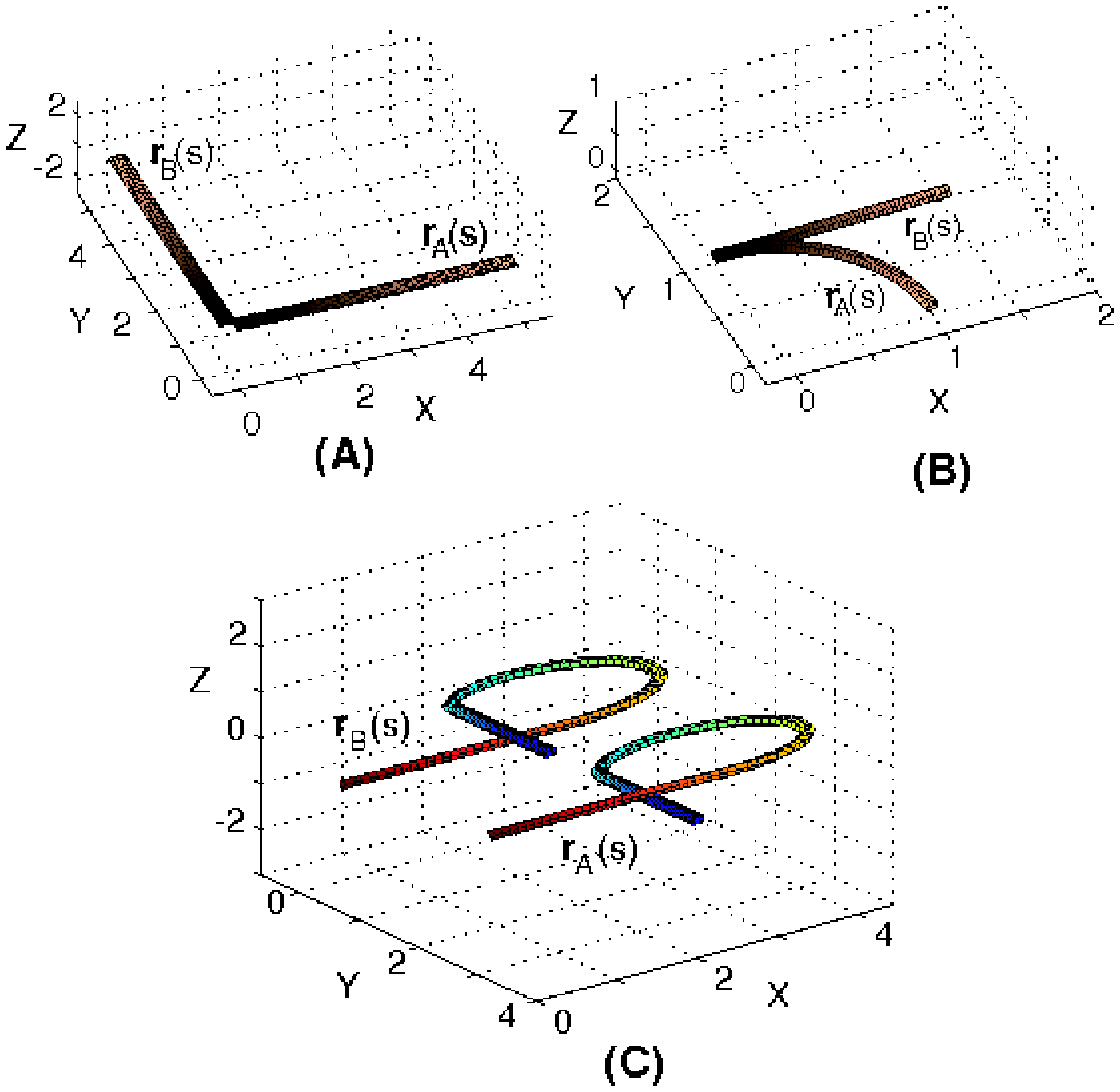}
\caption{\label{curvepairs}}
\end{figure}

\begin{figure}
\includegraphics[height=0.267\linewidth]{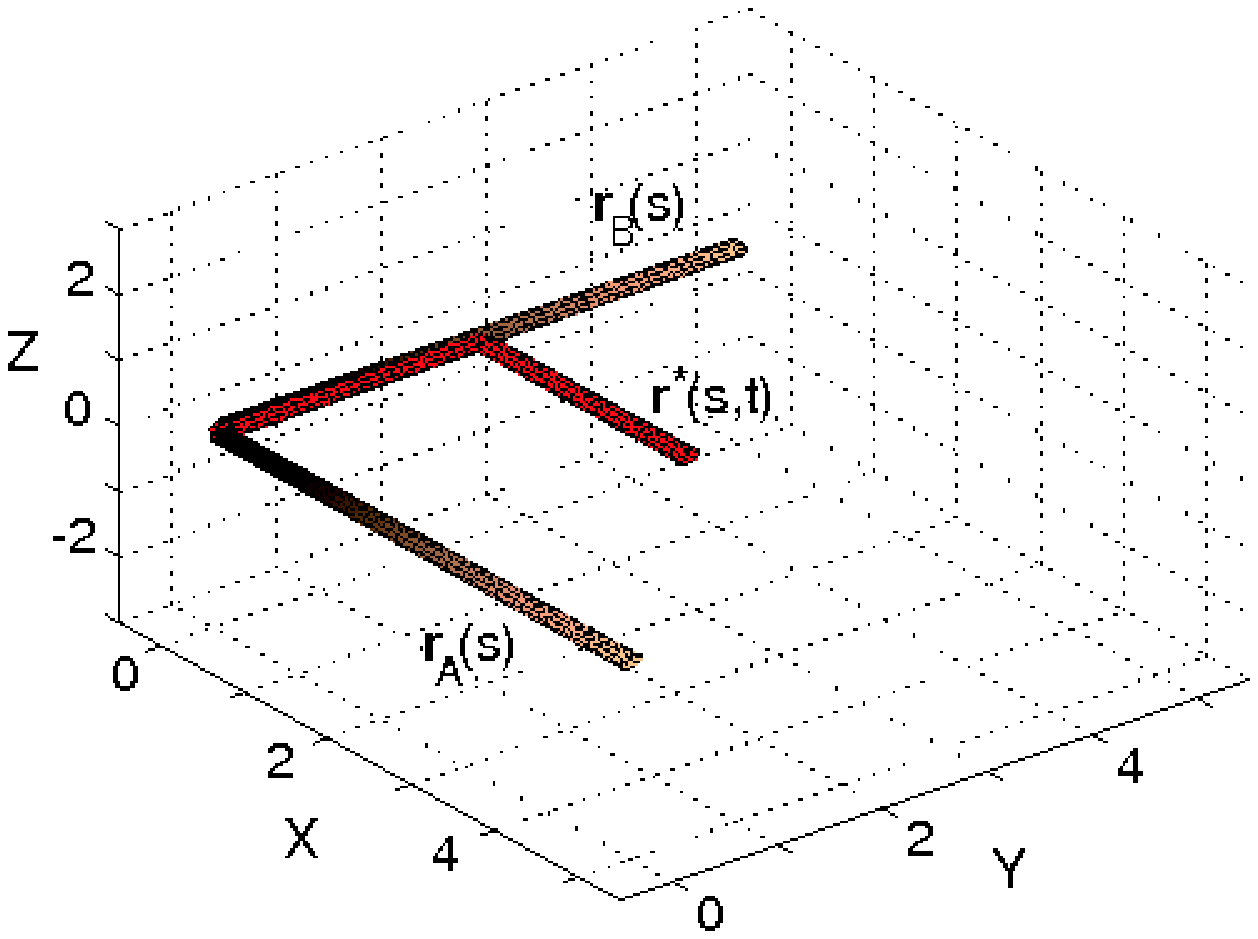}
\caption{\label{L-curve}}
\end{figure}

\begin{figure}
\includegraphics[height=0.314\linewidth]{link5.eps}
\caption{\label{fig1link}}
\end{figure}

\begin{figure}
\includegraphics[height=0.554\linewidth]{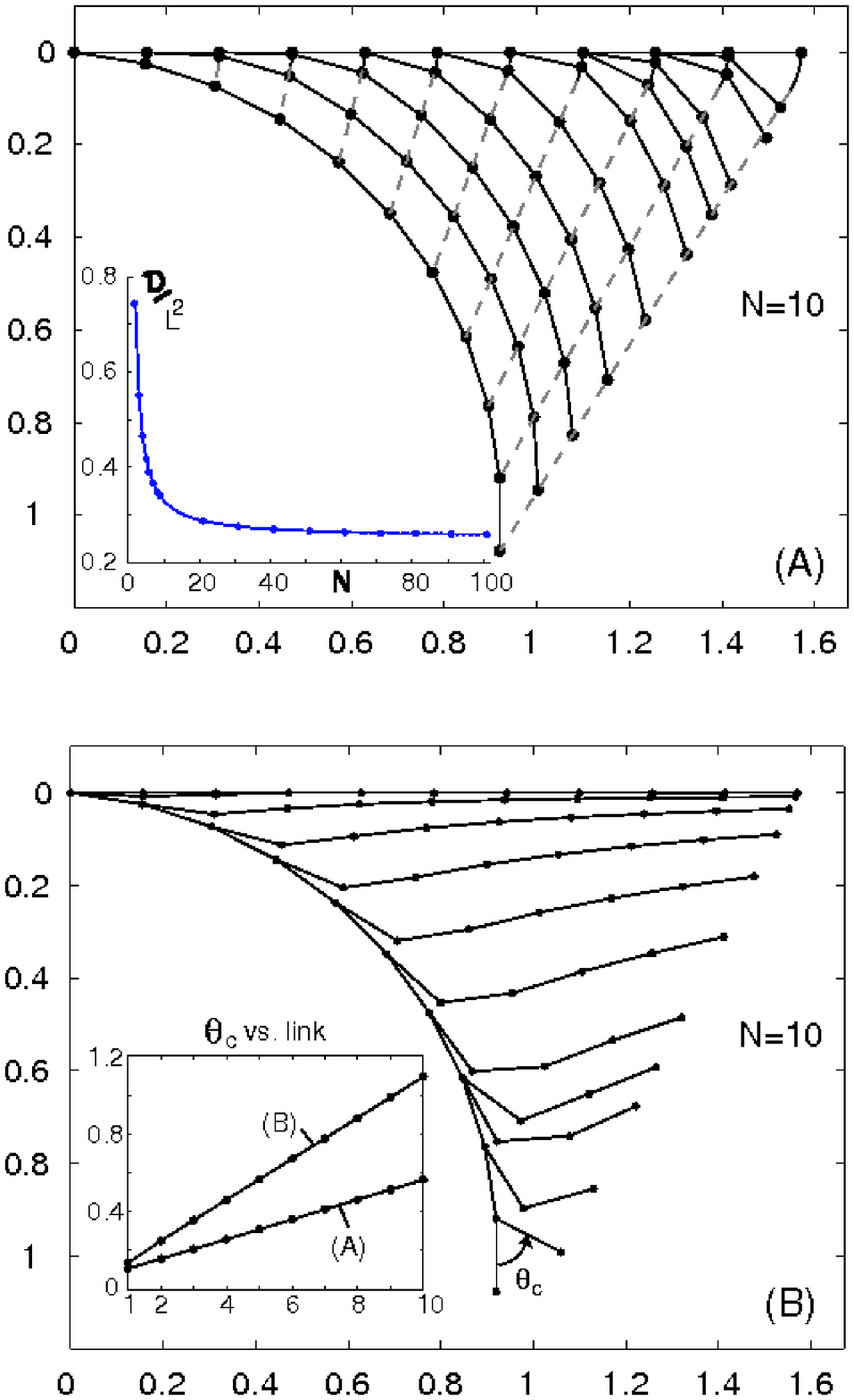}
\caption{\label{curvchaintot}}
\end{figure}

\renewcommand{\theequation}{\thesection.\arabic{equation}}
\renewcommand{\thesection}{\Alph{section}}
\setcounter{section}{0}

\end{document}